\documentclass[10pt]{article}
\usepackage{latexsym,graphicx,epstopdf}
\usepackage[left=2.5cm,top=2.5cm,right=2.5cm,bottom=1.5cm]{geometry}
\usepackage{textcomp}
\usepackage{amsmath}
\usepackage{amssymb}
\usepackage{graphicx}
\begin{document}
\begin{center}
\Large{\bf{ An FLRW interacting dark energy model of the Universe }} \\
\vspace{10mm}
\normalsize{Anirudh Pradhan$^1$, G. K. Goswami$^2$, A. Beesham$^3$, Archana Dixit$^4$ }\\
\vspace{10mm}
\normalsize{$^1$ Department of Mathematics, Institute of Applied Sciences and Humanities, G L A University,
    Mathura-281 406, Uttar Pradesh, India  \\
    \vspace{2mm}
    E-mail: pradhan.anirudh@gmail.com} \\
\vspace{2mm}
\normalsize{$^2$ Department of Mathematics, Kalyan P G College, Bhilai-490006, India}\\
\vspace{2mm}
\normalsize{Email: gk.goswami9@gmail.com} \\
\vspace{2mm}
\normalsize{$^3$ Department of Mathematical Sciences, University of Zululand, Kwa-Dlangezwa 3886, South Africa  \\
    \vspace{2mm}
    E-mail: beeshama@unizulu.ac.za} \\
    \vspace{2mm}
    \normalsize{$^4$ Department of Mathematics, Institute of Applied Sciences and Humanities, G L A University,
    Mathura-281 406, Uttar Pradesh, India  \\
    \vspace{2mm}
    E-mail: archana.dixit@gla.ac.in} \\
   New Astronomy 78 (2020) 101368 
\end{center}
%%%%%%%%%%%%%%%%%%%%%%%%%%%%%%%%%%%%%%%%%%%%%%%%%%%%%%%%%%%abstract%%%%%%%%%%%%%%%%%%%%%%%%%%%%%%%%%%%%%%%%%%%%%%%%%%%%%%%
\begin{abstract}
In this paper, we have presented an FLRW universe containing two-fluids (baryonic
and dark energy), with a  deceleration parameter (DP)
having a transition from past decelerating to the present accelerating universe. In
this model, dark energy (DE) interacts with dust to produce a new law for the
density. As per our  model, our universe is at present in a phantom phase after
passing through a quintessence phase in the past. The physical importance of the
two-fluid scenario is described in various aspects. The model is shown to
satisfy current observational constraints such as recent Planck results.
Various cosmological parameters relating to the history of the  universe have
been investigated.
\end{abstract}
%%%%%%%%%%%%%%%%%%%%%%%%%%%%%%%%%%%%%%%%%%%section1%%%%%%%%%%%%%%%%%%%%%%%%%%%%%%%%%%%%%%%%%%%%%%%%%%%%%%%%%%%%%%%%%%%%%%
 %%%%%%%%%%%%%%%%%%%%%%%%%%%%%%%%%%%%%%%%%%%section1%%%%%%%%%%%%%%%%%%%%%%%%%%%%%%%%%%%%%%%%%%%%%%%%%%%%%%%%%%%%%%%%%%%%%%
\section{Introduction}
A cosmological model must satisfy the basic cosmological principle (CP) which says that at any time the universe is
spatially homogeneous and isotropic. There is no privileged position in the universe. The Friedmann-Lemaitre-Robertson-Walker (FLRW) model
satisfies the CP. This  was manifest in  an expanding and decelerating universe filled with a  perfect fluid. However the latest findings
on observational grounds during the last three decades by various cosmological missions \cite{ref1}$-$\cite{ref17} confirm that
universe has an accelerating expansion at present. It is believed that there is a bizarre form of dark energy (DE) with negative
pressure  prevailing all over the universe which is responsible for the said acceleration. In $\Lambda$ CDM cosmology
\cite{ref18, ref19}, the  $\Lambda$-term is used as a candidate of DE with equation of state $ p_ \Lambda = -\rho _ \Lambda =
\frac{-\Lambda c^4}{8\pi G}$. However, the model suffers from, inter alia, fine tuning and cosmic coincidence problems \cite{ref20}.
Any acceptable cosmological model must explain the  accelerating  universe.\\

As of now,  many models and theories such as quintessence, phantom, $k$-essence, holographic DE models, $f(R)$ and $f(R,T)$
theories have been proposed to explain the acceleration in the universe. One may refer  to the review article \cite{ref18} for a brief
introduction to these models and theories. \\

 Of late, many authors \cite{ref21}$-$\cite{ref27} presented DE models in which the DE is considered in a conventional manner
 as a fluid with an EoS parameter $\omega_{de} = \frac{p_{de}}{\rho_{de}}$. It is assumed that our universe is filled with two
 types of perfect fluids of which one is a baryonic fluid (BF) which has positive pressure and creates deceleration in the universe.
 The other  is a DE fluid which has negative pressure and creates acceleration in the universe. Both fluids have different EoS parameters.
 The EoS for baryonic matter has been solved by cosmologists by providing the phases of the universe like stiff matter, radiation
 dominated and present dust dominated universe, but the determination of the EoS for DE is an important problem in observational
 cosmology at present. The present value of $\omega_{de}$  is observationally estimated nearly equal to $-1$. In the quintessence model,
 $-1 \leq\omega_{de}  < 0$ whereas in the phantom model $\omega_{de} \leq -1 $. Latest surveys \cite{ref28}$-$\cite{ref32}
 rule out the possibility of $\omega_{de} \ll -1$, but $\omega_{de}$ may be little less than $-1$. But we are facing fine tuning and 
 coincidence problems \cite{ref33}. So we need a dynamical DE with an effective EoS, $\omega_{(de)} = p_{(de)}/\rho_{(de)} < -1/3$. 
 The two types of surveys SDSS and WMAP \cite{ref9} and \cite{ref34} provide limits on $\omega_{(de)}$ as $-1.67 < \omega_{de} < -0.62$ 
 and $-1.33 < \omega_{de} < - 0.79$, respectively.\\

  It is worthwhile to mention here that various researchers  \cite{ref35} $-$ \cite{ref39} proposed that DE may  interact
 with BF, so they  have developed both types of interacting and non-interacting models of the universe. Recently it has been discovered that allowing
 and interaction between DE and dark matter(DM) offers an attractive alternative to the standard model of the cosmology \cite{ref40,ref41}.
 In these works the motivation to study interacting DE model arises from high energy physics. In recent work Risalti and Lusso \cite{ref42}
 and Riess {\it et al} \cite{ref43} stated that a rigid $\Lambda$ is ruled out by $~4\sigma$ and allowing for running vacuum favored phantom type
 DE ($\omega < -1$) and $\Lambda$ CDM is claimed to be ruled out by $4.4\sigma$ motivating the study of interacting DE models. Interacting DE
 models \cite{ref44} $-$ \cite{ref51} lead to the idea that DE and DM do not evolve separately but interact with each other non gravitationally
 (see recent review \cite{ref52} and references there in.). \\

  Motivated from above discussion, in this paper,  we have presented an FLRW universe containing two-fluids (baryonic
  and dark energy), with a  deceleration parameter (DP) having a transition from past decelerating to the present accelerating universe As per our  model,
   universe is at present in a phantom phase after passing through a quintessence phase in the past. The model is shown to satisfy current observational
    constraints such as Planck's latest observational results \cite{ref17}. Various cosmological parameters relating
 to the history of the universe have been investigated.\\

 Our paper is structured as follows: In Sec. $2$, we set the initial field equations. In Sec. $3$, we have described the results and physical properties 
 of interacting DE model. Finally, Sec. $4$ is devoted to our conclusions.

%%%%%%%%%%%%%%%%%%%%%%%%%%%%%%%%%%%%%%%%%%%%%%%%%%%%%%%%%%%%%%%%Section 2 %%%%%%%%%%%%%%%%%%%%%%%%%%%%%%%%%%%%%
\section{Field equations}

 The FLRW space-time (in units $c = 1$) is given by
\begin{equation}
\label{eq1}
ds{}^{2}=dt{}^{2}-a(t){}^{2}\left[\frac{dr{}^{2}}{(1+kr^{2})}+r^{2}({d\theta{}^{2}+sin{}^{2}\theta
    d\phi{}^{2}})\right],
\end{equation}
where $a(t)$ stands for the scale factor and $k$ is the curvature parameter.\\
 The  stress-energy tensor
 $ T_{ij} = T_{ij}(m)+T_{ij}(de),$ where $T_{ij}(m)=\left(\rho_m + p_m\right)u_{i}u_{j}-p_m g_{ij}$ and
 $T_{ij}(de)=\left(\rho_{de}+p_{de}\right)u_{i}u_{j}-p_{de} g_{ij}.$
  We assume that DE interacts with and transforms energy to baryonic matter.
We follow  arXiv:1905.10801 and 1906.00450 to get
Einstein field equations (EFEs)  for the FLRW metric (\ref{eq1}) are as follows.
\begin{equation}
\label{eq2}
H^2(1-\Omega_{de})=H^{2}_{0}\left[(\Omega_{m})_{0} \left(\frac{a_0}{a}\right)^{3(1-\sigma)}+(\Omega_{k})_{0}
\left(\frac{a_0}{a}\right)^{2}\right],
\end{equation}
and
\begin{equation}
\label{eq3}
2q = 1 + 3\omega_{de}\Omega_{de}+ 3\frac{H^2_0}{ H^2}
\omega_{k}(\Omega_{k})_{0} \left(\frac{a_0}{a}\right)^2 ,
\end{equation}
where symbols have their usual meanings.\\
 
\section{Results and disussions}

In the above, we have found two field equations (\ref{eq2}) and (\ref{eq3}) in five
unknown variables $a,~H,~q,\Omega_{de}$ and $ ~\omega_{de}$. Therefore, for
a complete solution, we need three more  relations involving these variables.
Many  researchers \cite{ref53} $-$ \cite{ref55} have considered constant DP which
is not valid from present observations. The DP $q$ may be taken as time
dependent as supported by many observations like SN Ia \cite{ref5,ref6,ref28}
and CMB anisotropies \cite{ref7,ref8}. From these observations, we observe that
$z < 0.5 $ for the present accelerated phase whereas $z > 0.5$ for the early
decelerating phase. Furthermore, the corrected red shift $z_{t} = 0.43 \pm 0.07$
by ($1\; \sigma$) c.1. \cite{ref8} from $z_{t} = 0.46 \pm 0.13$ at ($1\; \sigma$)
c.1. \cite{ref28} as of late found by the High-Z Supernova Search (HZSNS)
group. The Supernova Legacy Survey (SNLS) \cite{ref29}, and additionally the
one as of late incorporated by Knop {\it et al} \cite{ref33}, yields $z_{t} \sim 0.6
(1 \; \sigma)$ in better concurrence with the flat $\Lambda$CDM model ($z_{t} =
(2\Omega_{\Lambda}/\Omega_{m})^{\frac{1}{3}} - 1 \sim 0.66$). In this way, the
DP, which by theory is the rate with which the universe decelerates, must show
signature flipping \cite{ref56}$-$\cite{ref60}. From these discussions, $q$ may
not be taken as a constant, but it should be time-dependent. Recently, many
researchers \cite{ref35}$-$\cite{ref39} \& \cite{ref60}$-$\cite{ref63} have used the time-dependent
DP for solving various cosmological problems. So we consider $q$ as a linear
function of the Hubble function parameter which was earlier used by \cite{ref64}$-$\cite{ref66} in different context 
of cosmological models.
\begin{equation}\label{eq4}
 q  = \beta H + \alpha
\end{equation}
Here $\alpha$, and $\beta$ are arbitrary constants and unit of $\beta$ is Gyr as $H$ is expressed in $Gyr^{-1}$ and q is dimension 
less quantity.\\

From above equation, we have $ \frac{a \ddot{a}}{\dot{a}^{2}} + \beta
\frac{\dot{a}}{a} + \alpha = 0$, which on solving, yields
\begin{equation*}
 a = exp{\left[-\frac{(1 + \alpha)}{\beta}t -\frac{1}{(1 + \alpha)} + \frac{l}{\beta}\right]}, ~ ~ provided ~ \alpha \ne -1.
\end{equation*}
Here $l$ is a constant of integration. \\
From this, we calculate
\[
 \dot{a} = -\left(\frac{1 + \alpha}{\beta}\right)\exp{\left[-\left(\frac{1 + \alpha}{\beta}\right)t -
 \frac{1}{(1 + \alpha)} + \frac{l}{\beta}\right]},
\]
\begin{equation*}
 \ddot{a} = \left(\frac{1 + \alpha}{\beta}\right)^{2}\exp{\left[-\left(\frac{1 + \alpha}{\beta}\right)t -
\frac{1}{(1 + \alpha)} + \frac{l}{\beta}\right]}.
\end{equation*}
Putting above values in Eq. (\ref{eq4}), we obtain the DP value as $q = -1$.
Similarly we also observed that $q = -1$ for
$\alpha = 0$. \\
For $\alpha = -1$, we have to find another solution. In this case Eq. (\ref{eq12})
reduces to
\begin{equation*}
 q = -\frac{a \ddot{a}}{\dot{a}^{2}} = -1 + \beta H ,
\end{equation*}
which yields the following differential equation:
\begin{equation*}
 \frac{a\ddot{a}}{\dot{a}^{2}} + \beta \frac{\dot{a}}{a} -1 = 0.
\end{equation*}
The solution of above equation is found to be

\begin{equation}
\label{eq5}
 a = \exp{\left[\frac{1}{\beta}\sqrt{2\beta t + k}\right]},
\end{equation}

where $k$ is an integrating constant. \\

Since we are interested to study the cosmic decelerated-accelerated transit
universe, so we only consider the later case
for which $\alpha = -1$.\\

The derivation of Eq. (\ref{eq5}) can also be seen in \cite{ref65}. Now we
determine the constants  $\beta$ and $k$ on the basis of the latest observational
findings due to Planck \cite{ref17}. The values of the cosmological parameters at
present are as follows. $   (\Omega_{m})_{0} =  0.30 ~~ (\Omega_{k})_{0}= \pm
0.005,~~ (\omega_{de})_{0}=-1, (\Omega_{de})_{0}  =  0.70\pm0.005,~~
H_0=0.07 Gyr^{-1}~~ q_0   \simeq  -0.55,~~ t_{0}=13.72 Gyr.$ Eq. (\ref{eq13})
provides following  differential equation

\begin{equation}\label{eq6}
  (1+z) H_z = \beta H^2 = H(1+q) = \frac{\beta}{2 \beta t + k}
\end{equation}
where we have used $\frac{a_0}{a} = 1+z, \dot{z} = - (1+z)H$ and $ H_z = \frac{d
H}{dz}.$ From Eq. (\ref{eq6}) and the Planck results, we get the value of
constants $\beta$ and $k$ as
\begin{equation}\label{eq7}
 k = 27.6816~ Gyr{^2}, \beta = 6.42857~ Gyr
\end{equation}
Integrating  Eq. (\ref{eq6}), we get $H^{-1} = A - \beta log(1+z)$, where $A$ is constant of integration. As $H_{0}= 0.07$, $A= 100/7.$
So, we get following solution
\begin{equation}\label{eq8}
H= \frac{7}{100 - 45~ log(1 + z)}~ Gyr^{-1} ,~~q= \frac{45}{100\, -45 log (z+1)}-1.
\end{equation}

 %%%%%%%%%%%%%%%%%%%%%%%%%%%%%%%%%%%%%%%%%%%% Section 4 %%%%%%%%%%%%%%%%%%%%%%%%%%%%%%%%%%%%%%%%%%%%%%%%%%

 %%%%%%%%%%%%%%%%%%%%%%%%%%%%%%%%%%%%%%%%%%%%%%%%%%%%%%%%%%%%%%%%%%%%%%%%%%%%%%%%%%%%%%%%%%%%%%
\textbf{(i) Hubble function  $H$: }\\

 The determination of the two physical quantities  $H_{0}$ and  $q$ plays an important role to describe the evolution of the universe.
 $H_{0}$ provides us the rate of expansion of the universe which in turn helps in estimating the age of the universe, whereas the DP $q$
 describes the decelerating or accelerating  phases during the  evolution of the universe. From the last two decades,  many attempts
 have been made to estimate the value of the Hubble function 
 \cite{ref27}, \cite{ref67}$-$\cite{ref69}. For detailed discussions, readers are referred to Kumar \cite{ref27}.
 We present the following figures $1$, $2$ \& $3$  to illustrate the solution Eq. (\ref{eq8}) .
 %%%%%%%%%%%%%%%%%%%%%%%%%%%%%%%%%Figure 1,2,3 %%%%%%%%%%%%%%%%%%%%%%%%%%%%%%%%%%%%%%%%%%%%%%%%%%%%%
 %\begin{figure}[!h]
 \begin{figure}[ht]
    \centering
    \includegraphics[width=4cm,height=3cm,angle=0]{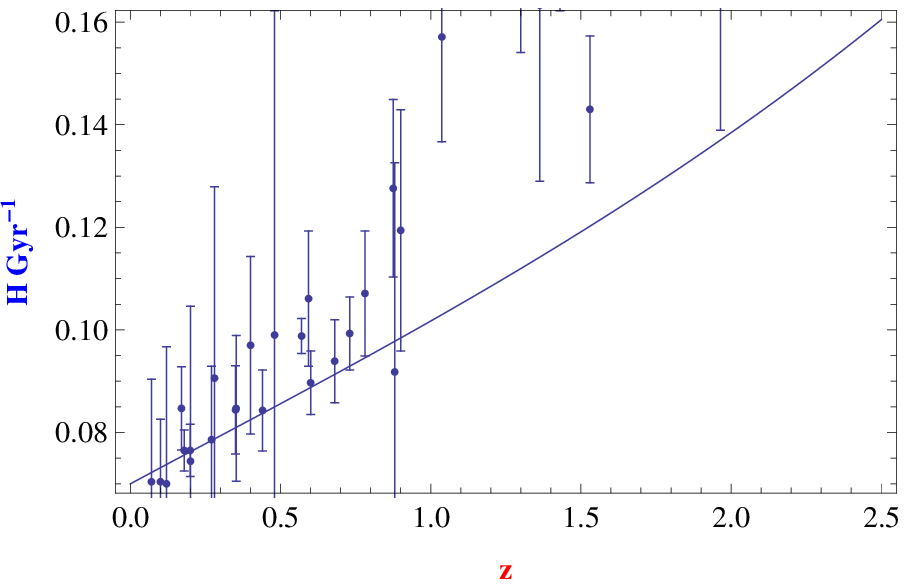}
     \includegraphics[width=4cm,height=3cm,angle=0]{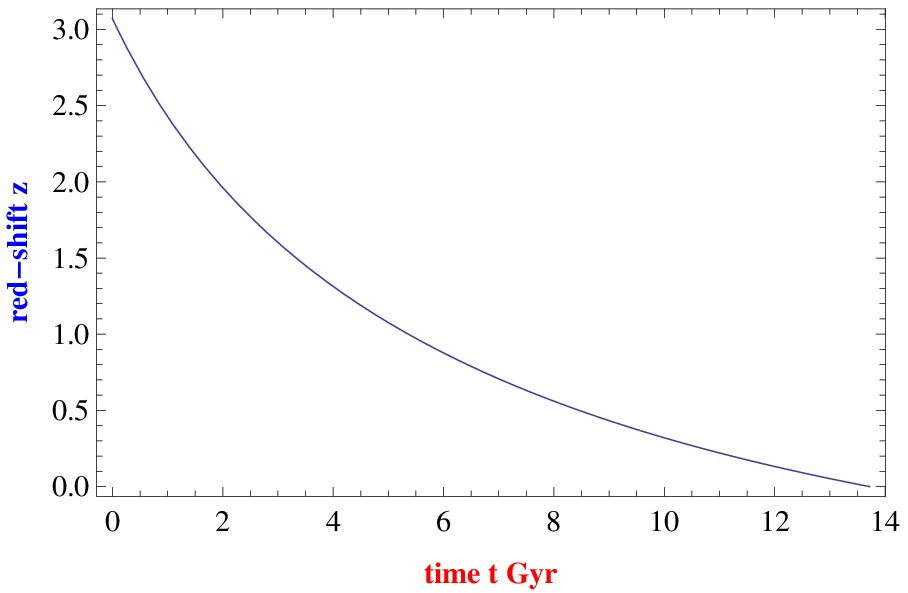}
     \includegraphics[width=4cm,height=3cm,angle=0]{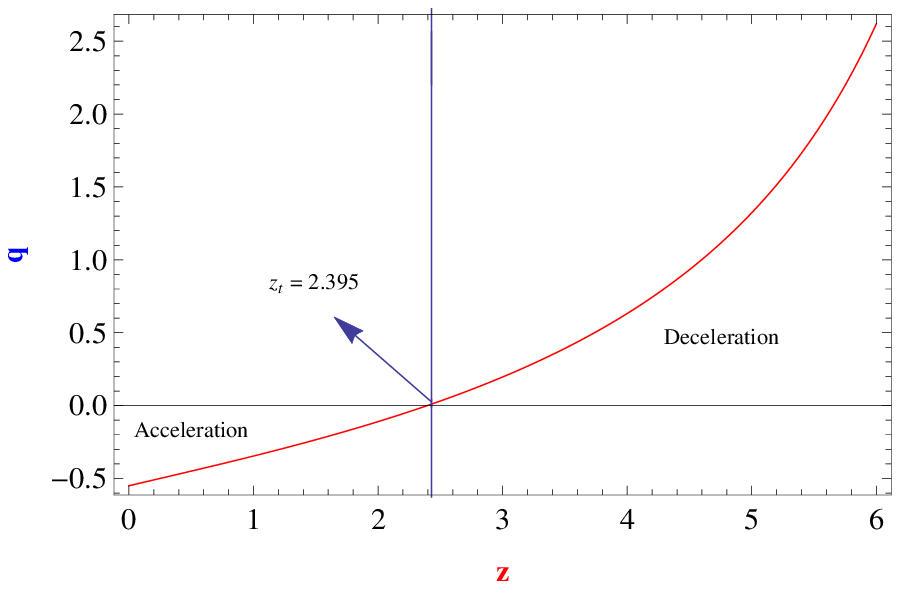}\\
     
  {\large Figures 1,2 and 3:  Plot of Hubble function  ($H$) versus red shift ($z$)(left),Variation of ($z$) versus  ($t$)(middle) and Variation of $q$ with $z$(right)}
 \end{figure}
 Various researchers \cite{ref15,ref16}, \cite{ref70} $-$\cite{ref76} have estimated values of the Hubble function at different
red-shifts using a differential age approach and galaxy clustering method [see
\cite{ref76} for list of 38 Hubble function parameters ].  We obtain $\chi^2$ from the following formula
   \begin{equation*}
     \chi^{2} = \sum _{i=1}^{i=38}[(Hth(i) - Hob(i))^2/\sigma(i)^2],
   \end{equation*}
   where $Hth$ (i)'s are theoretical values of Hubble function parameter as per Eq. (\ref{eq8}) and $\sigma(i)$'s are errors in 
   the observed values of $H(z)$. It comes to $\chi^{2}$= $33.22$ i.e. $87.43$ over $38$ data's, which shows best fit in theory and observation.
 From figure $1$, we observe that $H$ increases with
the increase of red shift. In this figure, cross signs are $31$ observed values of
the Hubble function $H_{ob}$ with corrections, whereas the linear curve is the
theoretical graph of the Hubble function  $H$ as per our model. Figure $2$ plots
the variation of red shift $z$ with time $t$, which shows that in the early universe the red shift was more than at present.\\

%%%%%%%%%%%%%%%%%%%%%%%%%%%%%%%%%%%%%%%%%%%%%%%%%%%%%%%%%%%%%%%%%%%%%%%%%%%%%%%%%%%%%%%%%
\textbf{(ii) Transition from deceleration to acceleration:}\\

Now we can obtain the DP `$q$' in term of red shift `$z$' by using
Eq.(\ref{eq8}). We present  figure $3$ to illustrate the solution. This describes the phase
variation of the universe from deceleration to acceleration.  We see that at present our universe is undergoing an accelerating phase.
It has begun at the transit red shift $z_t = 2.395$, i.e., at the time $T_t = 1.034$ Giga year. It was decelerating before time $T_t$\\

%%%%%%%%%%%%%%%%%%%%%%%%%%%%%%%%%%%%%%%%%%%%%%%%%%%%%%%%%%%%%%%%%%%%%%%%%%%%%%%%%%%%%%%%%
\textbf{(iii) DE Parameter $\Omega_{de}$ and EoS  $\omega_{de}$ }\\

  Now, from Eqs. (\ref{eq2}), (\ref{eq3}) and energy conservation equations, the density parameter $\Omega_{de}$ and EoS
parameter $\omega_{de}$ for DE are given by the following equations  and are solved numerically.
\begin{equation}
\label{eq9}
H^2 \Omega_{de} = H^2 - (\Omega_{m})_0 H^2_0 (1+z)^{3(1-\sigma)}
\end{equation}

\begin{equation}
\label{eq10}
\omega_{de}=\frac{H^2(2\alpha H +2\beta-1)}{3[H^2-H_0^2(\Omega_{m})_0(1+z)^{3(1-\sigma)}]}.
\end{equation}
where we have taken $(\Omega_{k})_0 = 0$ for the  present  spatially flat universe. We would take $ \sigma $ = $0.243$ for
numerical solutions to match with latest observations. We solve Eqs. (\ref{eq9}) and (\ref{eq10}) with the help of Eq. (\ref{eq8})
and present the following figures $3$ and $4$  to illustrate the solution. \\
%%%%%%%%%%% Figure 4 %%%%%%%%%%%%%%%%%%%%%%%%%%%
\begin{figure}[ht]
    \centering
    \includegraphics[width=6cm,height=5cm,angle=0]{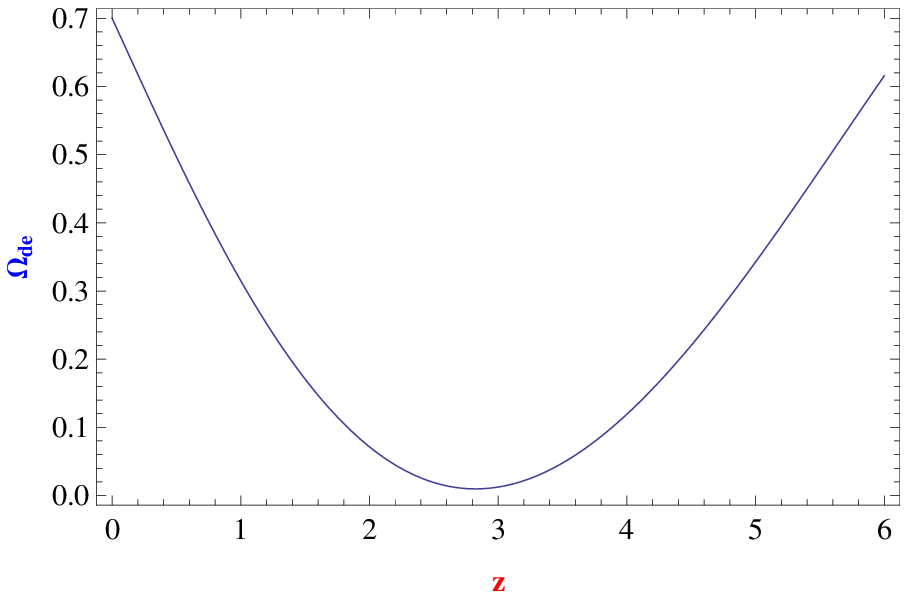}
    \includegraphics[width=6cm,height=5cm,angle=0]{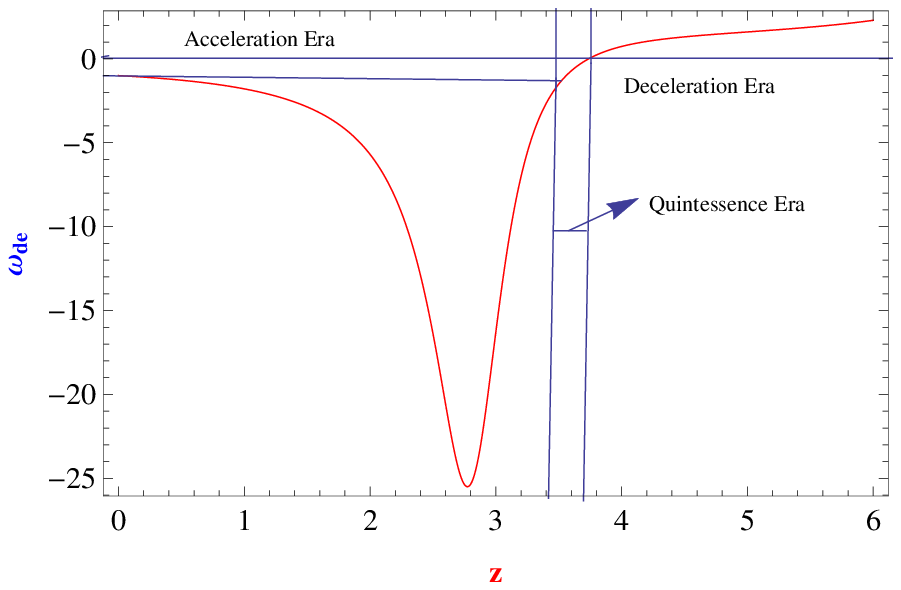}\\
   {\large  Figures 4 and 5: Plot of $\Omega_{de}$ versus red shift ($z$)(left) and plot of $\omega_{de}$ versus $z$(right). Phantom phase $(0 \leq z \leq 3.665)$, quintessence phase $ 3.665 \leq z \leq 3.74$ and  deceleration phase  $z \geq 3.74$}
\end{figure}
%%%%%%%%%%%%%%%%%%%%%%%%%%%%%%%%%%%%%%%%%%%%%%%%%%%%%%%%%%%%%%%%%%%%%%%%%%%%%%%%%%%%
%%%%%%%%%%%%%%%%%%%%%%%%%%%%%%%%%%%%%% Figure 5 %%%%%%%%%%%%%%%%%%%%%%%%%%%%%%%%%%

%%%%%%%%%%%%%%%%%%%%%%%%%%%%%%%%%%%%%%%%%%%%%%%%%%%%%%%%%%%%%%%%%%%%%%%%%%%%%%%%%%%%%%%%%%%%
Our model envisages that at present we are living in a phantom phase $\omega_{(de)}\leq -1$.  In the past at
$ z = 2.77~~ \omega_{(de)} = -25.4947$
was minimum, and then it started increasing. This phase remains for the period $(0 \leq z \leq 3.665)$. Our universe entered into a quintessence
phase at $ z = 3.665$ where $\omega_{de}$ comes up to $ -0.333123 $. As per our model, the
period for the quintessence phase is  the following
$$ 3.665 \leq z \leq 3.74.$$ DE favors deceleration at $z \geq 3.665$.
We look carefully Figs. $4$ and $5$ in context of Fig. $3$. In fact as per Fig. $3$,
the transition red shift is $z_{tr}$ = $2.395$. As we expressed in our explanation,
dark energy will begin its roll of opposing deceleration and favoring acceleration
during $0\leq z \leq 2.395$. Before, i.e., $z \geq 2.395$, universe is decelerating.
so dark energy as well as $\omega_{de}$ have no physical rolls. We may say
that the validity of Figs. $4$ and $5$ is only during the said tenure. During this DE
always increases with time.
 As per our model, the present ratio of DE is $0.7$. It decreases over the past, attains a  minimum value $\Omega_{de}=0.005$ at $z= 2.747$,
 and then it again increases with red shift.\\
%%%%%%%%%%%%%%%%%%%%%%%%%%%%%%%%%%%%%%%%%%%%%%%%%%%%%%%%%%%%%%%%%%%%%%%%%%%%%%%%%%%%%%%%%%%%%%%%%%%%

%%%%%%%%%%%%%%%%%%%%%%%%%%%%%%%%%%%%%%%%%%%%%%%%%%%%%%%%%%%%%%%Subsubsection 2.15  %%%%%%%%%%%%%%%%%%%%%%%%%%%%%%%%%%%%%%%%%%%%
\textbf{(iv) Distance modulus $\mu$ and Apparent Magnitude $m_{b}$:}\\

The distance modulus $\mu$ and apparent magnitude $m_b$ \cite{ref18}  are  derived as
\begin{equation}\label{eq11}
    \mu  =   m_{b}-M      =   5log_{10}\left(\frac{D_L}{Mpc}\right)+25 =  25+  5log_{10}\left[\frac{c(1+z)}{H_0}
    \int^z_0\frac{dz}{h(z)}\right]
 \end{equation}

\begin{equation}\label{eq12}
m_{b}=16.08+ 5log_{10}\left[\frac{1+z}{.026} \int^z_0\frac{dz}{h(z)}\right].
\end{equation}
We solve Eqs. (\ref{eq11})$-$ (\ref{eq12}) with the help of Eq. (\ref{eq8}). Our
theoretical results have been compared with SNe Ia related union $2.1$
compilation $581$ data \cite{ref14}, and the derived model was found to be in
good agreement with current observational constraints. The following figures $6$
depict the closeness of observational and theoretical results, thereby justifying
our model. In order to get quantitative closeness of theory and
observation, we obtain $\chi^2$ from the following formula
   \begin{equation*}
\chi^{2}= \sum _{i=1}^{Length SN1aData} \frac{\mu_{th} (i)-\mu_ {obs}(i))^2}{\sigma SN1a(i)^2}
    \end{equation*}
   where$ \mu_{th}$(i)'s are theoretical values of distance modulus as per Eq. (\ref{eq12}) and $\sigma SN1a(i)$'s are errors in the 
   observed values of $\mu$. It comes to $\chi^{2}$= 562.227 i.e. 96.7$\%$ over 581 data's, which shows best fit in theory and observation.

%%%%%%%%%%%%%%%%%%%%%%%%%%%%%%%%%%%%%%%%Figure 6 %%%%%%%%%%%%%%%%%%%%%%%%%%%%%%%%%%%%%%%%%%%%%%%%%%%%
\begin{figure}[!ht]
    \includegraphics[width=10cm,height=5cm,angle=0]{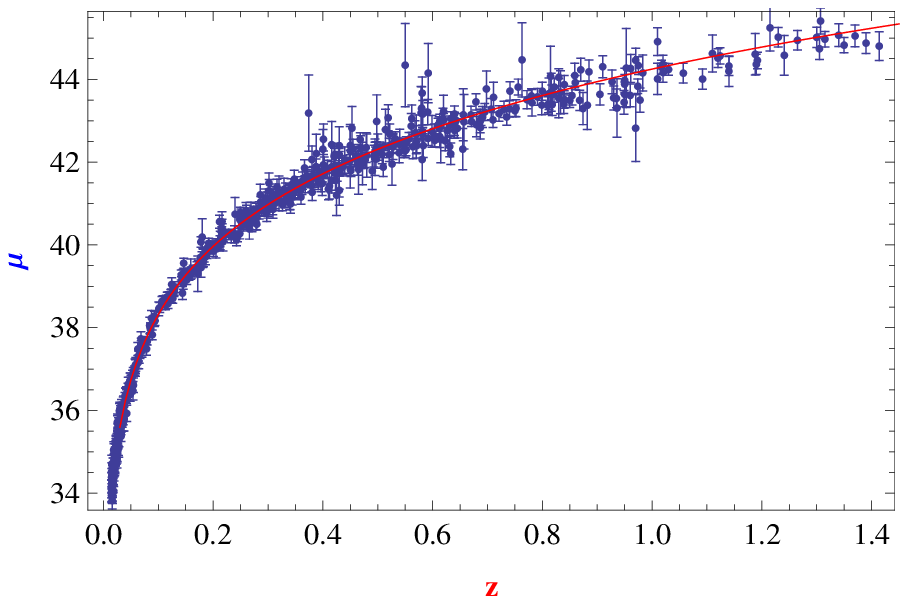}\\
{\large  Figure 6:    Plot of distance modulus ($\mu=M-m_b$) versus red shift ($z$). Crosses are  SNe Ia related union $2.1$ compilation 581 data's with possible corrections}
\end{figure}
%%%%%%%%%%%%%%%%%%%%%%%%%%%%%%%%%%%%%%%%%%%%%%%%%%%%%%%%%%%%Section5%%%%%%%%%%%%%%%%%%%%%%%%%%%%%%
\section{ Conclusion:}
In the present paper, we have presented an FLRW universe filled with two fluids
(baryonic and dark energy), by assuming a scale factor as a linear function of the
Hubble function . This results in a time-dependent DP having a transition from
past decelerating  to the present accelerating universe. The main findings of our
model are itemized point-wise as follows.
\begin{itemize}
    \item The expansion of the universe is governed by a  expansion law $ a(t)= (\beta H -1) = exp \frac{\sqrt{2\beta t+k}}{\beta}$, where
    $ \beta =  6.42857~ Gyr $ and $k = 27.6816~ Gyr^2$. This describes the transition from  deceleration to acceleration.
    \item  Our model is based on the recent observational findings due to the Planck results \cite{ref17}. The model agrees with
    present cosmological
    parameters.\\
    $(\Omega_{m})_{0}$= 0.30 $(\Omega_{k})_{0}= \pm0.005$, $(\omega_{de})_{0}=-1$, $(\Omega_{de})_{0}=0.70\pm0.005$, $H_0=0.07$
    Gy$^{-1}$, $q_0 = 0.055$  and present age $t_{0}=13.72$ Gy.
    \item  At present our universe is undergoing an accelerating phase. It has begun at the transit red shift $z_t = 2.395$, i.e.,
    at the time $T_t = 1.034$ Gigayear. It was decelerating before time $T_t$

    \item Our model has a variable EOS  $\omega_{de}$ for the DE density.  Our model envisages that at present we are living in the
    phantom phase $\omega_{(de)}\leq -1$.    In the past at
    $ z = 2.77~~ \omega_{(de)} = -25.4947$
    was minimum, then it started increasing. This phase remains for the period $(0 \leq z \leq 3.665)$. Our universe entered into
    a quintessence
    phase at $ z = 3.665$ where $\omega_{de}$ comes up to $ -0.333123 $. As per our model, the
    period for the quintessence phase is  the following
    $$ 3.665 \leq z \leq 3.74.$$ DE favors deceleration at $z \geq 3.665$.
    As per our model, the present ratio of DE is 0.7. It decreases over the past, attains a  minimum value $\Omega_{de}=0.005$ at
    $z= 2.747$, and then it again increases with red shift.
    \item The DE interacts with dust matter in our model, giving rise to a new density law for dust as
    $\rho_{m} = (\rho_{m}) _0\left (\frac{a_0}{a}\right)^{3(1-\sigma)}$,
    where $\sigma$ is a constant which has been assigned the value $0.243$ to match with observations.
\end{itemize}

In a nutshell, we believe that our study will pave the way to more research in future, in particular, in the area of the early
universe, inflation and galaxy formation, etc. The proposed hybrid expansion law may help in investigations of
hidden matter like dark matter, dark energy and black holes.
%%%%%%%%%%%%%%%%%%%%%%%%%%%%%%%%%%%%%%%%%%%%%%%%%%%%%%%%%%%%%%%%%%%%%%%%%%%%%%%%%%%%%%%%%%%%%%%%%%%%%%%%%%%%%%%%%%%%%%%%%
\section*{Acknowledgement} The authors (G. K. Goswami \& A. Pradhan) sincerely acknowledge the Inter-University Centre for Astronomy
and Astrophysics (IUCAA), Pune, India for providing facilities where part of this work was completed during a visit.

%%%%%%%%%%%%%%%%%%%%%%%%%%%%%%%%%%%%%%%%%%%%%%%%%%%%%%%%%%%%%%%%%%%%%%%%%%%%%%%%%%%%%%%%%%%%%%%%%%%%%%%%%%%%%%%%%%%%%%%%

\end{document}